# Activation measurements of an iodinated contrast media for online range verification in proton therapy

A. Espinosa-Rodriguez[a,b,*], V.V. Onecha[a,b,e], V.M. Nouvilas[a], S. Viñals i Onsès[c], P. Ibañez[a,b], S.España[a,b,d], D. Sánchez-Parcerisa[a,b], J.M. Udías[a,b], L.M. Fraile[a,b]

a Grupo de Física Nuclear, EMFTEL & IPARCOS, Universidad Complutense de Madrid, CEI Moncloa, Madrid, Spain
b Instituto de Investigación del Hospital Clínico San Carlos (IdISSC), Ciudad Universitaria, Madrid, Spain
c Centro de Microanálisis de Materiales, CMAM-UAM, Madrid, Spain
d Centro Nacional de Investigaciones Cardiovasculares (CNIC), Madrid, Spain
e Department of Radiation Oncology, Massachusetts General Hospital and Harvard Medical School, Boston, MA, United States
*Correspondence: anespi04@ucm.es

## ABSTRACT

The use of contrast agents has previously been proposed as a novel method to increase the activation close to the Bragg peak, aiming to improve the quality of proton range monitoring in vivo. In a recent work, we demonstrated the feasibility of $^{127}$I for online verification, thanks to its high cross-section (200 mbarn at 10 MeV) and low energy production threshold for $^{127m}$Xe. The spectroscopy-based method relies on identifying two specific delayed γ lines (at 124.6 and 172.5 keV), which can be analyzed via single-photon emission computed tomography (SPECT). In this work, we present a proof-of-principle study to investigate proton activation in a commercial iodinated contrast media (ICM) for radiology. Five measurements were conducted at different proton energies (6-10 MeV), equivalent to the last millimeters of clinical proton range. Activation in the ICM was measured with four LaBr$_3$(Ce) scintillators. The contribution from iodine was separated from the activation of the solvent, yielding excellent agreement with previously reported data. These results demonstrate the potential of this technique and pave the way for further testing in clinically relevant scenarios.

## 1. INTRODUCTION

Proton therapy (PT) is the most widely practiced form of particle therapy, with over 120 centers around the world [PTCOG]. The rise in popularity of this radiotherapy technique during the last decades can be ascribed to the superior dose deposition pattern of protons inside the human body, which results in much better tumor targeting [Baumann2016]. Unlike conventional radiotherapy with photons, most of the dose is deposited in a well-defined narrow area, the so-called Bragg peak, after which it sharply decreases to zero [Wilson, 1946].

However, clinical implementation of this technique is currently hampered by uncertainties in the Bragg peak position, which are caused by the translation of Hounsfield units to proton stopping power, the dose calculation techniques, inter- or intra-fractional anatomical changes during treatment, or target delineation, among others [Paganetti, 2012]. Consequently, the proton range in the patient bears a few mm of uncertainty, which reduces the overall efficacy of PT and prevents this technique from attaining its full potential [Paganetti et al., 2021].

Nuclear reactions induced by protons on human tissue produce secondary radiation or other emissions that can be correlated to the proton range. This is essentially the basis of the main in vivo range verification techniques, which rely on the detection and imaging of positron emitters, implemented as PET (Positron Emission Tomography), prompt gamma (PG) rays from short-lived excited nuclei, MRI imaging or ionoacoustic waves [Knopf and Lomax, 2013; Takayanagi et al., 2020, Freijo et al., 2021].

Regarding PET, real-time feedback of the dose deposition can only be achieved with short-lived isotopes (with half-lives ranging from ns to sec) as this minimizes the influence of biological wash-out and the need to reposition the patient during treatment [Ferrero et al., 2019]. To date, the in-beam PET modality, in which the scanner is integrated into the treatment room, is a relevant technique in this area inasmuch as PET nuclear imaging is already well-established in the medical field [Ferrero et al., 2018]. However, the high radiation background levels that are present during the treatment and the non-conventional PET detector geometry needed challenge the in-beam modality [Pausch et al., 2020]. On the other hand, although PG radiation is produced instantaneously, its clinical implementation also experiences major drawbacks due to the high energy (up to a few MeV) of the emitted photons, the strong gamma-ray flux, and again the high amount of background radiation [Jeyasugiththan et al., 2021; Martins et al., 2021]. Furthermore, measurement of the activation of naturally existing isotopes adds extra complexity because these reactions exhibit high production thresholds or low integrated cross-sections, resulting in poor count statistics near the end of the proton range [España et al., 2011].

One method that has been proposed to address this is the use of fiducial markers, but this is an invasive procedure which requires the insertion of one or several metal markers, along the beam path, in the tumor or close to this area. Furthermore, this technique is restricted to assessing the residual range of the beam relative to the marker position but does not provide 3D information about the beam's position. [Burbadge et al., 2021; Nemallapudi et al., 2022; Kasanda et al., 2023a; Kasanda et al., 2023b]. Similarly to what is already done in medical imaging, an alternative solution could be the administration of contrast agents. The term was already introduced in this field to describe elements that can enhance the proton activation within the tumor volume yielding a favourable spatial distribution, very close to the Bragg Peak [España et al., 2021, Rodriguez et al., 2021, España et al., 2022, Onecha et al., 2023]. Some isotopes that have been proposed as potential contrast agents for range verification in protontherapy are $^{68}$Zn, $^{18}$O, or $^{127}$I [España et al., 2021; Rodriguez et al., 2021, Onecha et al., 2023, Fraile et al., 2016; Ibañez-Moragues et al., 2023;].

Of all of them, $^{127}$I is a very promising candidate for real-time dose verification since the half-life of $^{127m}$Xe produced by proton activation is very short ($T_{1/2}$ = 69.2 s). The radioactive decay of $^{127m}$Xe is characterized by the emission of two characteristic gamma rays at 124.6 keV (69 % relative intensity) and 172.5 keV (38 %) which are well suited for spectroscopic measurements and do not overlap with the gamma emission that again characterizes the activation of natural isotopes in human tissue. Therefore, a tomographic image of the region can be acquired shortly after the treatment if a sufficient number of gammas are detected by a SPECT system. Alongside PET, SPECT is one of the primary methods currently employed in diagnostic medicine. For instance, the use of radioiodine in the form of the $^{123}$I and $^{131}$I isotopes is a well-established procedure in nuclear imaging, to measure its volume and activity distribution in the human body, utilizing a SPECT scanner [Morphis et al., 2021]. The use of SPECT for online monitoring has been also proposed in Boron Neutron Capture Therapy (BNCT) and Proton Boron Fusion Therapy (PBFT), imaging the 478 and 719 keV prompt gamma rays emitted in the two boron reactions [Kobayashi et al., 2000, Shin et al., 2016], as well for the 718 keV prompt gamma emissions in carbon range verification [Parajuli et al., 2021].

Clinical implementation of proton range verification using an iodine contrast agent requires accurate knowledge of the production cross-section for proton induced reaction, especially at low energies, to correlate the measured activity range to the clinical dose range. Recently, we reported [Rodriguez et al., 2021] new cross-section data for the $^{127}$I(p,n)$^{127m}$Xe reaction and significant activity was observed in very close proximity (0.3 mm) to the Bragg Peak. The maximum cross-section value measured was 200 mb at 9.45 MeV, which makes this signal readily distinguishable from the background radiation.

On the other hand, iodinated contrast media (ICM) are commonly used in many radiographic procedures and interventional treatments because they are low-cost and already approved for human administration. Given the potential of this technique for in-vivo range verification in proton therapy, in this work, we aim at studying the proton activation on a clinical ICM. This approach also allows for testing the sensitivity of this technique at lower iodine concentrations and against a water background. Measurements were performed at different proton energies in the low energy range at the Centre for Micro Analysis of Materials [Redondo-Cubero et al., 2021]. The activity was monitored in real-time, and the production yields were validated against previous measurements on a solid (pressed powder) CsI target.

## 2. MATERIALS AND METHODS

The experimental procedure is very similar to the one described in our previous work [Rodriguez et al., 2021]. Irradiations were performed at the Centre for Micro Analysis of Materials (CMAM-UAM). The external microbeam line was used to extract proton beams at energies of 6, 7, 8, 9 and 10 MeV and intensities ranging from 2.0 to 10.5 nA. Current stability was monitored online, looking at the detector rates and a posteriori, via the irradiation logbooks of the accelerator. Measurements of gamma radiation were conducted online. The length of the irradiation period was selected to ensure reaching at least 50% of the saturation activity in the sample, whereas decay of the activation products was recorded until the activity dropped below the background level.

### 2.1. Experimental setup

Spectroscopic measurements were performed with four LaBr$_3$(Ce) detectors positioned around the sample at angles of 115º, 40º, 15º and 90º relative to the beam direction (see Figure 1a). All elements were fixed with 3D printed holders to a solid desk and the full setup was aligned using the piezoelectric XYZ positioning stage and the beamline laser, which points towards the beam direction. The efficiency of the four detectors was obtained using a $^{152}$Eu calibrated source, placed at the same location as the target.

For this study, we selected a commercial iodine-based contrast media (Iomeron-400; 400 mg/mL, Bracco Imaging SpA. Milan, Italy). Iomeron contains 816.6 mg of iomeprol ($C_{17}H_{22}I_3N_3O_8$) per mL, trometamol, hydrochloric acid and water for injection. Prior to each irradiation, 450 μL of the solution were transferred to a 600 μL Eppendorf tube. An example is illustrated in Figure 1b. The tube was positioned horizontally and opened on the holder, facing the beam exit at a distance of 25(2) cm. Activation was recorded during both beam-on and beam-off periods, and the detector signals were processed using a high-resolution digitizer (CAEN DT 5751) with 10 bits vertical resolution and a sampling rate of 1 GS/s.

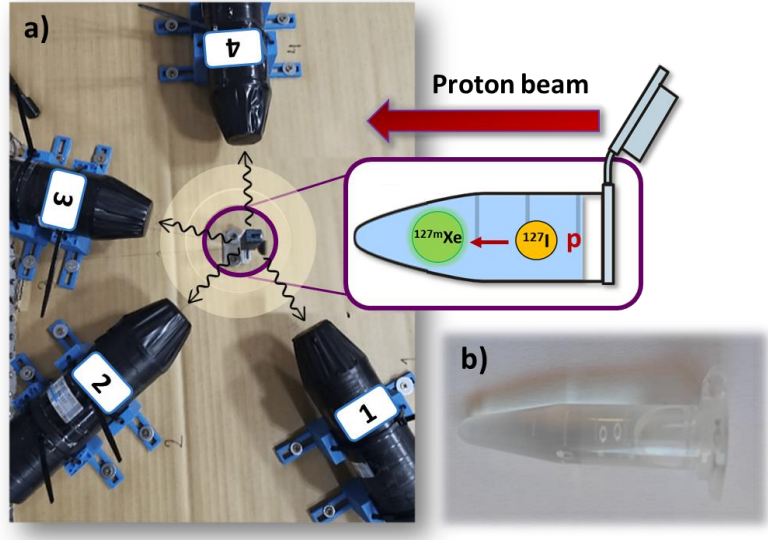

Figure 1. a) Experimental setup used for measuring proton activation in the iodinated contrast media. b) Example of the target: 600 µL Eppendorf tube filled with 450 µL of the contrast agent.

## 2.2. Data analysis

Proton range in the contrast agent was estimated using SRIM [Ziegler et al., 2010] to have a maximum value of 980(40) µm at the highest proton energy of 10 MeV. Thick target yields (number of produced reactions per incoming projectile) in the sample were experimentally determined using the well-known activation formula [Krasnov, 1974]:

$$Y = \frac{A_0}{\phi(1-e^{\lambda t_{irr}})\varepsilon_\gamma I_\gamma} \quad (1)$$

where $A_0$ represents the activity of the $^{127m}$Xe isotope at the end of the irradiation, $\phi$ denotes the proton flux (s$^{-1}$), $t_{irr}$ stands for the total irradiation time (s) and $\varepsilon_\gamma$ and $I_\gamma$ represent the detector system efficiency and branching ratio at each gamma emission, respectively. To determine the activity at the end of the irradiation, the two gamma peaks (124.6 and 172.5 keV) were fitted to a Gaussian function and a linear background. Then, events in a ROI of width ±3σ were processed in equally spaced time intervals to obtain the corresponding measured activity in the sample. The value of $A_0$ was obtained by fitting the decay curves to a decaying exponential plus a constant background.

In this analysis, we have considered the production of $^{13}$N via the $^{16}$O(p,X)$^{13}$N reaction in the water solvent, which has a low energy threshold of 5.55 MeV. The β+ emitters contribute via Compton interactions to the residual background of the 172.5 keV gamma emission in the spectra. Since the half-life of the $^{13}$N is about ten orders of magnitude greater than that of $^{127m}$Xe (see Table 1), the two isotopes can be readily separated, and the decay curves have been fitted to a sum of exponentials including both contributions.

Finally, beam energy and position of the sample were estimated from the nominal values of the accelerator, considering the distance between the beam exit and the thickness of the Kapton window, using the formula in [Sanchez-Parcerisa et al., 2021]:

$$E_s(z, E_0) = [(E_0 - sE_0^q)^p - \frac{z}{r}]^{1/p} \qquad (2)$$

Where r,p,s and q are fit parameters which take the following values: r=2.08 cm · MeV$^{-1.75}$; p=1.75; s=0.265 MeV$^{1.73}$; and q=–0.73. $E_0$ and $E_s$ are expressed in MeV and correspond to the nominal and the incident energy. Finally, z is the air gap distance, expressed in cm.

*Table 1. Decay properties of the nuclides considered in the analysis of the activation in the sample.*

| Reaction | Half-life | Energy threshold (MeV) | Energy (keV) | BR (%) |
|---|---|---|---|---|
| $^{127}$I(p,n)$^{127m}$Xe | 69.2(9) s | 1.74 | 124.6/172.5 | 69/38(3) |
| $^{16}$O(p,X)$^{13}$N | 9.965(4) min | 5.55 | 511 | 199.6 |

## 2.3. Uncertainty analysis and correction factors.

The total uncertainty in the yields reported in this work is calculated by combining statistical and systematic errors. We have considered the same contributions explored in our past experimental work, including uncertainty in the initial activity (1-28%), beam intensity (1-5%), irradiation time, and detector efficiency (1%). Since only the surface of the ICM is activated, significant absorption is observed for the two gamma emissions in the detectors placed at 15, 40 and 90⁰. These losses were corrected by examining the ratio of the net peak areas at 124.6 and 172.5 keV relative to the detector placed at 115⁰. At the lowest proton energy considered in these measurements, self-absorption in the sample is about 60% higher for the 124.6 keV emission compared to the 172.5 keV emission. Finally, a second correction for gamma self-absorption (5% contribution) was also included to account for the 2.5 mm thick plastic holder surrounding the Eppendorf tube during the full set of measurements. Both contributions were added yielding relative uncertainties ranging from 4 to 48%, at the minimum proton incident energy of 6 MeV, which constitutes the largest contribution to the total experimental uncertainty reported in this work. Finally, we combined all of these contributions as nuisance parameters following a random distribution. The uncertainty in the beam energy is very low (< 0.5%) and has been estimated using error propagation laws from the uncertainties in the beam-source distance of 25±2 mm and the intrinsic uncertainty of the nominal energy.

## 3. RESULTS

### 3.1. Energy spectrum

Figure 2a shows the resulting energy spectrum of the gamma events recorded in the four detectors during the beam-off period. A background spectrum, obtained after irradiation (9 MeV) of an empty Eppendorf tube is shown in blue. In contrast, the red spectrum illustrates a similar spectrum obtained after irradiating the tube filled with 450 µL of the ICM. Both spectra are normalized to the irradiation length. Activation on the ICM produces the two characteristic gamma emissions at 124.6 and 172.5 keV, which dominate the spectrum and are not present in the absence of the media. Moreover, an increase in the net peak areas of the 511 keV is also observed due to the activation of the sample solvent. Both spectra exhibit two peaks at 674 and 981 keV, which are attributed to the activation of the beamline. A non-normalized zoom in the 100-200 keV region of this spectrum is displayed in Figure 2b, with the gamma events shown separately in the four detectors. Self-attenuation is clearly observed in three of the four detectors, and is particularly significant for the third one, placed behind the sample.

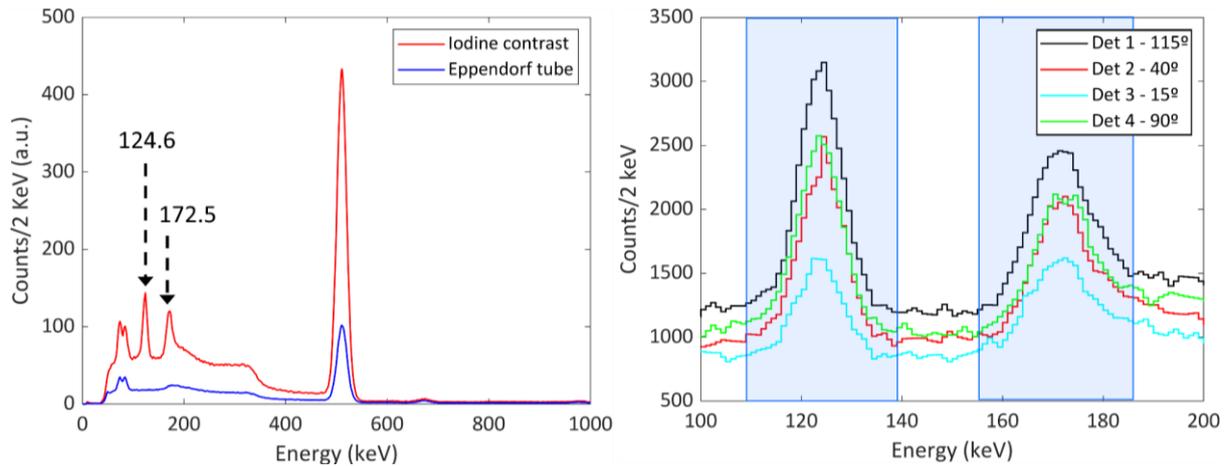

Figure 2. a) Gamma ray energy spectra recorded during beam-off periods and normalized to the length of the irradiation, following activation of an empty (blue) and ICM-filled (red) targets at 9 MeV. b) A zoom into the 100-200 keV region showing the total number of events of the activated ICM registered with the four detectors.

## 3.2. Decay curves

The time spectrum for the events registered in the 124.6 and 172.5 keV peaks is illustrated in Figures 3a and 3b after activation of the ICM-filled sample at 9 MeV. The shaded areas in Figure 2b indicate the events used in the analysis of the two gamma emissions. The origin (t=0) of the time axis depicted in both graphs corresponds to the start of the beam-off period. For the events in the 172.5 keV peak, the production of the $^{127m}$Xe is considered alongside the production of $^{13}$N to account for backscattered $\beta^+$ events resulting from the activation of $^{16}$O in the contrast (see section 2.2). In this case, the decay of $^{127m}$Xe is dominant during the first 150 seconds of the measurement. A constant term arising from background activity is also displayed in both figures.

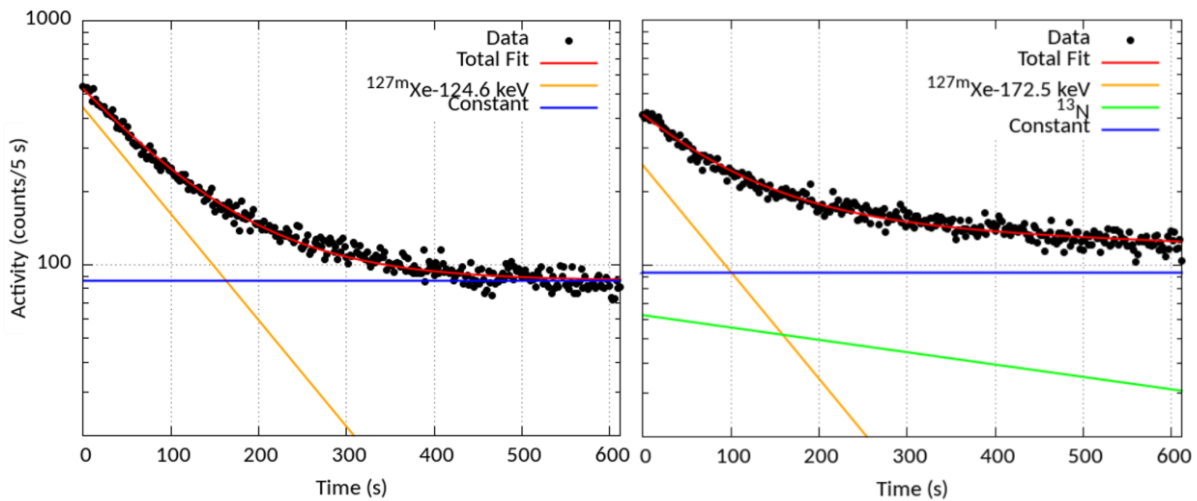

Figure 3. Decay curves and fitting of the activation products after irradiation at 9 MeV during the beam-off period: a) Events detected in the 124.6 keV peak and b) Events detected in the 172.5 keV peak along with the contribution of the $^{13}$N from the Compton interactions in this energy window.

## 3.3. Activation of the Iodinated Contrast Media.

Thick target yields (TTY) in the ICM for the $^{127}$I(p, n)$^{127m}$Xe reaction are presented in Figure 4 as a function of incident proton energy. Table 2 lists the corresponding values, which have been calculated using Eq.1 and combining the initial activity of the two gamma emissions at 125 and 173 keV, weighted by their branching ratios. Figure 4 also includes the previous data measured for this quantity using a CsI target [Rodriguez et al., 2021]. To account for the differences in atomic density between both samples, these measurements were scaled down by a factor of 5.51, corresponding to the ratio ($\rho_{at,\,CsI}$ : $\rho_{at,\,ICM}$). Excellent agreement is found between this work and our previous measurements. Only at the lower incident proton energy of 5.75 MeV, we observe a slight discrepancy. However, all of our past measurements lie within the uncertainty band due to the large uncertainty of this value (68%), which stems mostly from the low activation in the sample and large self-absorption in the media.

| Proton energy (MeV) | Intensity (nA) | TTY/10$^{-6}$ (reactions/projectile) |
|---|---|---|
| 9.83 ± 0.01 | 1.9 ± 0.1 | 39 ± 2 |
| 8.81 ± 0.01 | 4.4 ± 0.1 | 19.2 ± 0.9 |
| 7.80 ± 0.01 | 2.5 ± 0.1 | 8 ± 2 |
| 6.77 ± 0.02 | 5.0 ± 0.1 | 2.4 ± 0.4 |
| 5.75 ± 0.02 | 10.0 ± 0.1 | 0.5 ± 0.3 |

*Table 2. List of experimental thick target yields measured in this work.*

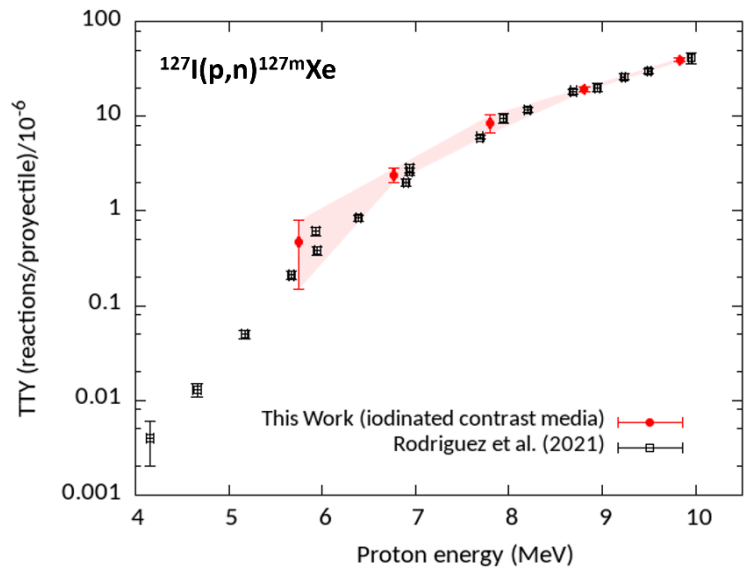

*Figure 4. Production yields measured in the iodinated contrast media as a function of the proton incident energy. Previous measurements in a CsI target are also included, scaled down by the ratio of iodine atomic densities [Rodriguez et al., 2021].*

## 4. DISCUSSION

In this work and in the framework of the PRONTO-CM project [PRONTO], we have further continued with the experimental campaign of measurements to assess the feasibility of using iodine as a contrast agent for proton range verification [Rodriguez et al., 2021]. The activation has been successfully measured online on a well-established clinical iodinated contrast media, and the results showed good agreement with our previously reported data, validating the efficacy of the proposed method. The measured activity was essentially free of background signal interference from other natural isotopes activated on the ICM in this energy region, and only a small Compton contribution from these PET emitters was observed in the 172.6 keV emission, yielding a univocal $^{127m}$Xe activation during the treatment.

Iodinated contrast agents are commonly employed in imaging diagnostics, particularly in computed tomography (CT) and angiography, and millions of radiological examinations are conducted annually with this type of intravascular contrast [Singh and Daftary, 2008]. Adverse effects related to the use of these substances are very low, showing a total incidence of 3.13%, and being the most common contrast-induced nephropathy (CIN), which is relevant only in patients with pre-existing abnormal renal function [Katayama et al., 1990].

Iodine concentration of the contrast agent selected for this study was 400 mg/mL. This is the maximum value of available formulations in the market, which typically range from iodine concentrations of 150 to 400 mg/mL. In patients, the average iodine administration is 0.5 to 1.5 mL per 1 kg of body weight, corresponding to an average of 30 g of iodine for a 75 Kg weight person. [Zanardo et al., 2018, Kessler et al., 2014]. However, in children, this quantity can be increased up to 4 mg/kg. Total iodine retention has been determined in different tumour volumes after a typical dosage 160 mL of Iomeron 400, the one employed here, reporting an average value of 1.94±0.12 mg/mL [Obeid et al., 2014]. This concentration remained stable in the volume for 20 minutes, providing sufficient time for in-vivo range verification with $^{127m}$Xe.

For a tumor size of 8.5 cm$^3$, this would result in a total iodine loading of 16.5 mg [Obeid et al., 2014]. This quantity is approximately one-tenth of the total iodine amount used in the measurements presented here. However, given the low toxicity of iodine, various strategies can be used to increase the local concentration in this area and should be explored to ensure maximal concentration of $^{127}$I in the tumor and surrounding tissues. Recently, several studies have suggested the possibility of using iodinated nanoparticles or liposomes to perform CT studies, which can be targeted to specific areas to enhance local iodine concentration and half-life [Hainfeld et al., 2019; Badea et al., 2019; Ghaghada et al., 2016; Zhang et al., 2021; Hsu et al., 2020] or to perform in-vivo imaging [Wallyn et al., 2018].

For clinical application, a high detection efficiency SPECT system and fast image reconstruction algorithm should be envisaged. In this work, a set of four LaBr$_3$ detectors was employed. However, a full-ring SPECT system or an array of detectors could be utilized to increase the detection efficiency. Compact SPECT systems have been already reported for BNCP applications, where low quantities of boron (~ppm) must be detected in a short time period [Abbene et al., 2022; Caracciolo et al., 2023]. In real conditions, the use of an appropriate collimator to perform imaging should be also considered. Given the spectroscopic similarities to $^{123}$I, a radioisotope typically used in SPECT, imaging in this case can be conducted using either a low-energy high-resolution (LEHR) or a medium-energy (ME) collimator [Morphis et al., 2021]. Therefore, it can be anticipated that any of these commercially available systems could be adapted to meet the requirements of the proposed method.

The use of a iodinated contrast media might contribute to an increasing signal near the end of the proton range, but its clinical feasibility remains to be investigated. Future studies will be necessary to determine the sensitivity limit based on iodine concentration, also involving more realistic scenarios, including homogeneous and non-homogeneous phantoms.

## 5. CONCLUSIONS & OUTLOOK

The activation of $^{127}$I, a potential contrast agent for online range verification in proton therapy, has been investigated in a commercial iodinated contrast media (ICM). Results show unambiguous $^{127m}$Xe activation, which can be clearly distinguished from other elements present in the sample. The obtained experimental data is in excellent agreement with previously published data on a solid thick target. For 8-MeV protons, accuracy better than 5% can be achieved, which corresponds to the distal end of a clinical beam. The observed results support the feasibility of ICM as a non-invasive procedure for enhancing the quality of proton monitoring through SPECT imaging. Further studies will be performed to assess the capabilities of the proposed system in a more realistic clinical scenario.

## ACKNOWLEDGEMENTS

This work was funded by Comunidad de Madrid, Spain under projects PRONTO-CM B2017/BMD-3888 and ASAP-CM S2022/BMD-7434. We acknowledge support by Spanish MCIN/AEI/10.13039/501100011033 under grants TED2021-130592B-I00 PROTOTWIN and PID2021-126998OB-I00. The authors acknowledge the support from The Centro de Microanálisis de Materiales (CMAM)—Universidad Autónoma de Madrid, for the beam time, and its technical staff for their contribution to the operation of the accelerator. This is a contribution for the Moncloa Campus of International Excellence, ''Grupo de Física Nuclear-UCM'', Ref. 910059.